\documentclass[prd,twocolumn,showpcs,amsmath,amssymb,nofootinbib,preprintnumbers,balancelastpage]{revtex4}
\pdfoutput=1
\usepackage{epsfig}
\usepackage{amsmath}
\usepackage{bm}
\usepackage{times}
\usepackage{graphicx}
\usepackage{color}
\usepackage{slashed}
\usepackage{graphicx}
\usepackage{amsmath}

\def\bea{\begin{eqnarray}}
\def\eea{\end{eqnarray}}
\def\ba{\begin{eqnarray}}
\def\ea{\end{eqnarray}}
\def\be{\begin{equation}}
\def\ee{\end{equation}}

\begin{document}
\preprint{CALT 68-2898}

\title{Simplified models with baryon number violation but no proton decay}

\author{Jonathan M. Arnold, Bartosz Fornal and Mark B. Wise\\
\textit{California Institute of Technology, Pasadena, CA 91125, USA}\\
}
\date{\today}

\begin{abstract}
We enumerate the simplest models that  have baryon number violation at the classical level but do not give rise to proton decay. These models have  scalar fields in two representations of $SU(3)\times SU(2)\times U(1)$ and violate
baryon number by two units. Some of the models give rise to $n \bar{n}$ (neutron-antineutron) oscillations, while some also violate lepton number by two units. We discuss the range of scalar masses for which  $n \bar{n}$ oscillations are measurable in the next generation of experiments. We give a brief overview of the phenomenology of these models and then focus on one of them for a more quantitative discussion of $n\bar n$ oscillations, the generation of the cosmological baryon number, the electric dipole moment of the neutron, and $K^0$-$\bar{K}^0$ mixing.
\vspace{11mm}
\end{abstract}

\maketitle
\bigskip

\section{Introduction}
The standard model has non-perturbative violation of baryon number ($B$). This source of baryon number non-conservation also violates lepton number ($L$), however, it conserves baryon number minus lepton number ($B-L$). The violation of baryon number by non-perturbative weak interactions is important at high temperatures in the early universe but it has negligible impact on laboratory experiments that search for baryon number violation and we neglect it in this paper.
If we add massive right-handed neutrinos that have a Majorana mass term and Yukawa couple to the standard model left-handed neutrinos, then lepton number is violated by two units, $|\Delta L|=2$, at tree-level in the standard model.

Motivated by Grand Unified Theories (GUT) there has been an ongoing search for proton decay (and bound neutron decay). The limits on  possible decay modes are very strong. For example, the lower limit on the partial mean lifetime for the mode $p \rightarrow e^+ \pi^0$ is $ 8.2\times10^{33} {\ \rm yrs}$ \cite{Nishino:2009aa}.  For proton decay to occur there must be interactions that violate baryon number by one unit and lepton number by an odd number of units.  See Ref. \cite{Nath:2006ut} for a review of proton decay in extensions of the standard model.

There are models where baryon number is violated  but proton (and bound neutron) decay does not occur. This paper is devoted to finding the simplest  models of this type and  discussing some of their phenomenology.  We include all renormalizable interactions allowed by the $SU(3) \times SU(2) \times U(1)$ gauge symmetry.  In addition to standard model fields they have scalar fields  $X_{1,2}$  that couple to quark bilinear terms  or lepton bilinear terms.  Baryon number violation either occurs through trilinear scalar interactions of the type  (i) $X_2 X_1 X_1$ or  quartic scalar terms of the type (ii) $X_2X_1X_1X_1$. The cubic scalar interaction in  (i) is similar in structure to renormalizable terms in the superpotential that give rise to baryon number violation in supersymmetric extensions of the standard model.  However, in our case the operator is dimension three and is in the scalar potential.  Assuming no right-handed neutrinos there are four models of type (i) where each of the $X$'s couples to quark bilinears and has baryon number $-2/3$. Hence in this case the $X$'s are either color ${\bf 3}$ or ${ \bf \bar 6}$. There are also five models of type (ii) where  $X_1$ is a color ${\bf 3}$ or ${ \bf \bar 6}$ with baryon number $-2/3$ that couples to quark bilinears  and $X_2$ is a  color singlet with lepton number $-2$ that couples to lepton bilinears.

We analyze one of the models in more detail. In that model the $SU(3)\times SU(2)\times U(1)$ quantum numbers of the  new colored scalars are $X_{1} \in \left(\bar{6}, 1, -1/3\right)$  and $X_{2} \in \left(\bar{6}, 1, 2/3 \right)$. The $n\bar n$ oscillation frequency is calculated using the vacuum insertion approximation for the required hadronic matrix element and  lattice QCD results. For coupling constants equal to unity and all mass parameters equal, the present absence of observed $\bar n n$ oscillations provides a  lower limit on the scalar masses of around $500~{\rm TeV}$. If we consider the  limit  $M_{1}\ll M_{2}$ then for $M_{1}= 5{\ \rm TeV}$  the next generation of $n \bar n$ oscillation experiments will be sensitive to $M_{2}$ masses at the GUT scale.

There are three models that have $n \bar{n}$ mixing at tree-level without proton decay.  In these models, constraints on flavor changing neutral currents and the electric dipole moment (edm) of the neutron make it unlikely that, if $n \bar{n}$ oscillations are observable, one of the scalar masses approaches the GUT scale.

In the next section we enumerate the models and discuss their basic features.  The phenomenology of one of the models is discussed in more detail in section III. Some concluding remarks are given in section IV.

\section{The models}
\begin{figure}[t]
\centering \hspace{5mm}
\includegraphics[scale=.65]{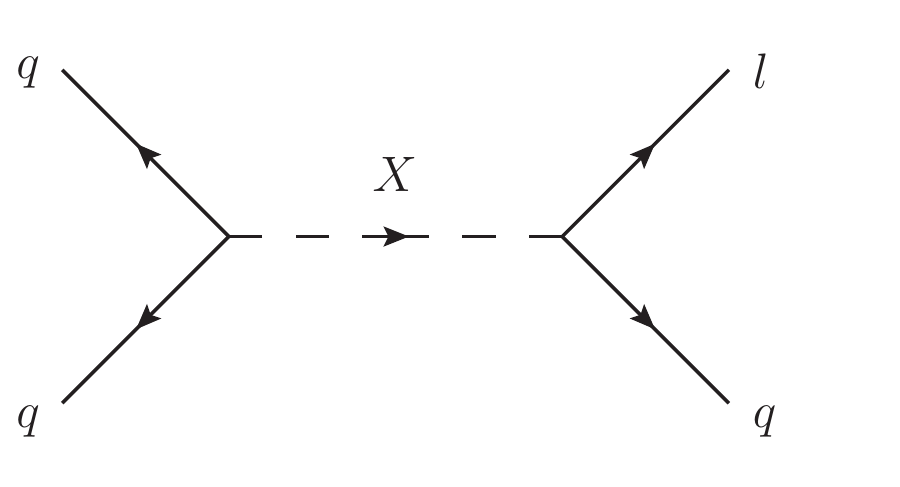}
\caption{$\Delta B = 1$ and $\Delta L = 1$ scalar exchange.}
\label{fig1}
\end{figure}

We are looking for the simplest models which violate baryon number but don't induce proton decay. We don't impose any global symmetries. Hence, all local renormalizable interactions permitted by Lorentz and gauge invariance are assumed to be present. We begin by considering renormalizable scalar couplings with all possible standard model fermion bilinears.  A similar philosophy can be used to construct models involving proton decay \cite{Barr:2012xb} or baryon number violating interactions in general \cite{Bowes:1996xy,Baldes:2011mh}. We first eliminate any scalars which produce proton decay via tree-level scalar exchange as in Fig.\ \ref{fig1}.
\begin{figure}[t]
\centering \hspace{5mm}
\includegraphics[scale=.65]{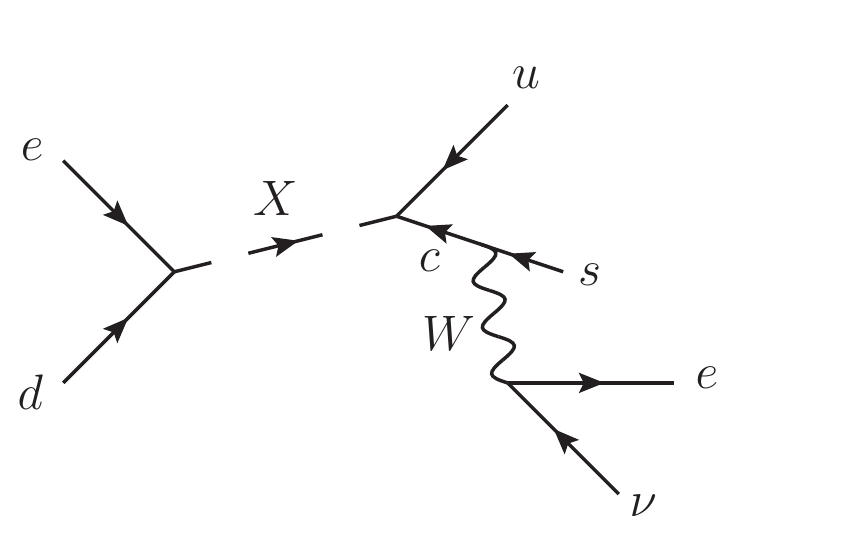}
\caption{Feynman diagram that contributes to tree-level $p \rightarrow K^+ e^+ e^- \bar{\nu}$ from $(3,1,-4/3)$ scalar exchange.}
\label{fig1c}
\end{figure}
In particular, this eliminates the scalars with $SU(3) \times SU(2) \times U(1)$ quantum numbers $(3,1,-1/3)$,
$(3, 3, -1/3)$, and $(3, 1, -4/3)$. Note that in the case of $(3,1,-1/3)$ we need an additional $W$-boson exchange to get proton decay (Fig.\ \ref{fig1c}) since the Yukawa coupling to right-handed charge 2/3 quarks is antisymmetric. The remaining possible scalar representations and Yukawa couplings are listed in Table \ref{table1}.  We have assumed there are no right-handed neutrinos ($\nu_R$) in the theory.
\begin{table}[b]
\begin{center}
    \begin{tabular}{| c | c |}
    \hline
       operator & \ \ \ \ \ $ SU(3) \times SU(2) \times U(1)$\ \ \ \ \   \\ \hline\hline
    $\ \ XQQ, Xud \ \ $ & $\left(\bar{6},1, -1/3\right)$  \\ \hline
    $XQQ$ & $\left(\bar{6},3, -1/3\right)$  \\ \hline
    $Xdd$ & $\left(3,1, 2/3\right)$, $\left(\bar{6},1, 2/3\right)$   \\ \hline
    $Xuu$ & $\left(\bar{6},1, -4/3\right)$  \\ \hline
    $X\bar{Q}e$ & $\left(3,2, 7/6\right)$  \\ \hline
    $X\bar{L}u$ & $\left(\bar{3},2, -7/6\right)$  \\ \hline
    $X\bar{L}d$ & $\left(\bar{3},2, -1/6\right)$  \\ \hline
    $XLL$ & $\left(1,1, 1\right)$, $\left(1,3, 1\right)$  \\ \hline
    $Xee$ & $\left(1,1, 2\right)$  \\
    \hline
    \end{tabular}
\end{center}
\caption{\footnotesize{Possible interaction terms between the scalars and fermion bilinears along with the corresponding quantum numbers.}}
\label{table1}
\end{table}

None of these scalars induces baryon number violation on their own, so we consider minimal models with the requirement that only two unique sets of scalar quantum numbers from Table \ref{table1} are included, though a given set of quantum numbers may come with multiple scalars.

\begin{figure}[t!]
\centering \hspace{5mm}
\includegraphics[scale=.61]{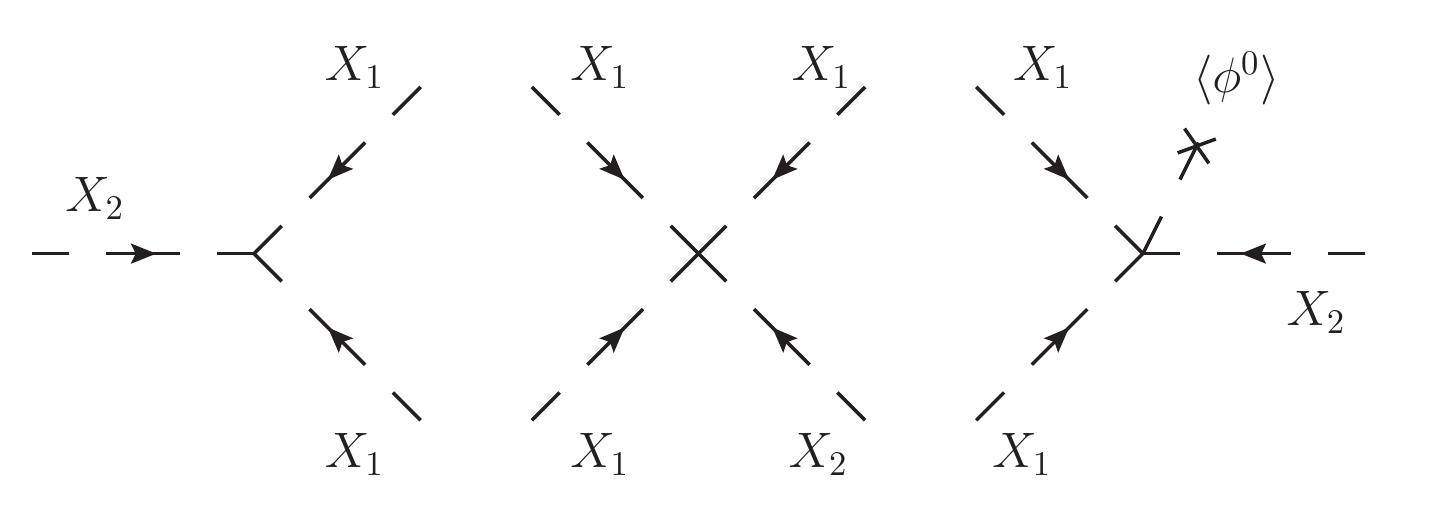}
\caption{Scalar interactions which may generate baryon number violation.}
\label{fig2}
\end{figure}

Baryon number violation will arise from terms in the scalar potential, so we need to take into account just the models whose scalar quantum numbers are compatible in the sense that they allow scalar interactions that violate baryon number. For scalars coupling to standard model fermion bilinears there are three types of scalar interactions which may violate baryon number: 3-scalar $X_1 X_1 X_2$, 4-scalar $X_1 X_1 X_1 X_2$, and 3-scalar with a Higgs $X_1 X_1 X_1 H$ or $X_1 X_1 X_2 H$, where the Higgs gets a vacuum expectation value (vev) (Fig.\ \ref{fig2}).

\begin{figure}[t!]
\centering \hspace{5mm}
\includegraphics[scale=.65]{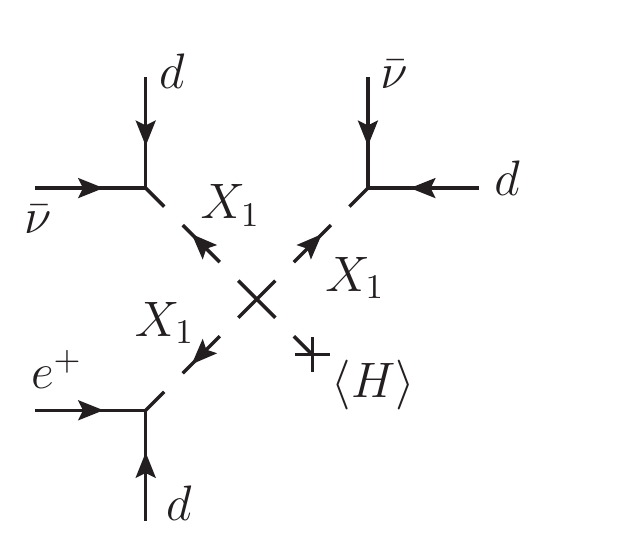}
\caption{Interaction which leads to proton decay, $p \rightarrow \pi^+ \pi^+ e^- \nu \nu$, for $X_1 \in (\bar{3}, 2, -1/6)$.}
\label{fig2b}
\end{figure}

Actually, the simplest possible model violating baryon number through the interaction $X_1 X_1 X_1 H$ includes just one new scalar $(\bar{3}, 2, -1/6)$, but it gives proton decay via $p \rightarrow \pi^+ \pi^+ e^- \nu \nu$ (Fig.\ \ref{fig2b}). The other two baryon number violating models with an interaction term $X_1 X_1 X_2 H$ are: $X_1^* \in (3, 1, -1/3)$, $X_{2} \in (\bar{3}, 2, -7/6)$ and $X_1 \in (3, 1, -1/3)$, $X_{2}^* \in (\bar{3}, 2, -1/6)$. As argued earlier, such quantum numbers for $X_1$ also induce tree-level proton decay, so we disregard them.

We now consider models with a 3-scalar interaction $X_1 X_1 X_2$. A straightforward analysis shows that there are only four models which generate baryon number violation via a 3-scalar interaction without proton decay. We enumerate them and give the corresponding Lagrangians below. All of these models give rise to processes with $\Delta B = 2$ and $\Delta L = 0$, but only the first three models contribute to $n\bar{n}$ oscillations at tree-level due to the symmetry properties of the Yukawas.  Note that a choice of normalization for the sextet given by,
\bea
(X^{\alpha \beta}) = \left(
      \begin{array}{ccc}
        \tilde{X}^{11} & \tilde{X}^{12}/\sqrt{2} & \tilde{X}^{13}/\sqrt{2} \\
        \tilde{X}^{12}/\sqrt{2} & \tilde{X}^{22} & \tilde{X}^{23}/\sqrt{2} \\
        \tilde{X}^{13}/\sqrt{2} & \tilde{X}^{23}/\sqrt{2} & \tilde{X}^{33} \\
      \end{array}
    \right)\
\eea
leads to canonically normalized kinetic terms for the color singlet elements $\tilde{X}^{\alpha \beta}$ and the usual form of the scalar propagator with symmetrized color indices.  Unless otherwise stated, we will be using 2-component spinor notation.  Parentheses indicate contraction of 2-component spinor indices to form a Lorentz singlet.
\\
\\
\noindent $\textbf{Model 1.\ \ \ }$ $X_{1} \in \left(\bar{6}, 1, -1/3\right)$, \ $X_{2} \in \left(\bar{6}, 1, 2/3\right)$
\bea\label{Lag1}
\mathcal{L} & = &  -\ g_{1}^{a b} X_{1}^{\alpha \beta} \left(Q_{L \alpha}^a \epsilon \,Q_{L \beta}^b\right)- {g}_{2}^{a b}  X_{2}^{\alpha \beta} (d_{R \alpha}^a d_{R \beta}^b)\nonumber\\
& & \hspace{-8mm} - \ {g}_{1}^{\prime a b}  X_{1}^{\alpha \beta} (u_{R \alpha}^a d_{R \beta}^b) + \lambda X_{1}^{\alpha \alpha'} X_{1}^{\beta \beta'} X_2^{\gamma \gamma'}\epsilon_{\alpha \beta \gamma} \epsilon_{\alpha'\beta'\gamma'}\
\eea
\noindent
By virtue of the symmetric color structure of the ${\bf \bar{6}}$ representation and the antisymmetric weak structure of the $QQ$ bilinear in the first term, $g_1$ must be antisymmetric in flavor.  However, this antisymmetry is not retained upon rotation into the mass eigenstate basis. Similarly, $g_2$ must be symmetric because of the symmetric color structure in the second term.  In this case, the symmetry character of $g_2$ will be retained upon rotation into the mass eigenstate basis because it involves quarks of the same charge.  Therefore, the interaction involving the Yukawa coupling $g_2$ gives rise to (and is thus constrained by) $K^0$-$\bar{K}^0$ mixing through tree-level $X_2$ exchange. The coupling $g_1^{\prime}$ has no particular flavor symmetry.
\\ \\
\noindent$\textbf{Model 2.\ \ \ }$  $X_{1} \in \left(\bar{6}, 3, -1/3\right)$, \ $X_{2} \in \left(\bar{6}, 1, 2/3\right)$
\bea
\mathcal{L} &\!\! =\!\! &  -\ g_{1}^{a b} X_{1}^{\alpha \beta A} (Q_{L \alpha}^a \epsilon \,\tau^A \,Q_{L \beta}^b)  - {g}_{2}^{a b}  X_{2}^{\alpha \beta} (d_{R \alpha}^a d_{R \beta}^b) \nonumber\\
& & + \ \lambda X_{1}^{\alpha \alpha' A} X_{1}^{\beta \beta' A} X_2^{\gamma \gamma'}\epsilon_{\alpha \beta \gamma} \epsilon_{\alpha'\beta'\gamma'}\
\eea
\noindent
Here the matrix $\epsilon \, \tau^A$ is symmetric.  Because the first and second terms have symmetric color structures, $g_1$ and $g_2$ must be symmetric in flavor.  The weak triplet $X_1$ has components which introduce both $K^0$-$\bar{K}^0$ and $D^0$-$\bar{D}^0$ mixing.  As in model 1, the interaction involving $g_2$ will introduce $K^0$-$\bar{K}^0$ mixing via $X_2$ exchange.
\\ \\
\noindent$\textbf{Model 3.\ \ \ }$  $X_{1} \in \left(\bar{6}, 1, 2/3\right)$, \ $X_{2} \in \left(\bar{6}, 1, -4/3\right)$
\bea
\mathcal{L} & = &  - \ g_{1}^{a b} X_{1}^{\alpha \beta} (d_{R \alpha}^a d_{R \beta}^b)  - \ {g}_{2}^{a b}  X_{2}^{\alpha \beta} (u_{R \alpha}^a u_{R \beta}^b)\nonumber\\
& & + \ \lambda X_{1}^{\alpha \alpha'} X_{1}^{\beta \beta'} X_{2}^{\gamma \gamma'}\epsilon_{\alpha \beta \gamma} \epsilon_{\alpha'\beta'\gamma'}\
\eea
\noindent
Both terms have symmetric color structures and no weak structure, so $g_1$ and $g_2$ must be symmetric in flavor.  In this model, the interactions involving $g_1$ and $g_2$ each have the potential to introduce neutral meson-antimeson mixing. For example, the $g_1$ interaction will induce $K^0$-$\bar{K}^0$ mixing while $g_2$ will induce $D^0$-$\bar{D}^0$ mixing.
\\ \\
\noindent$\textbf{Model 4.\ \ \ }$ $X_{1} \in \left(3, 1, 2/3\right)$,  \ $X_{2} \in \left(\bar{6}, 1, -4/3\right)$
\bea
\mathcal{L} & = &  -\ g_{1}^{a b} X_{1\alpha} \left(d_{R \beta}^a \,d_{R \gamma}^b\right)\epsilon^{\alpha\beta\gamma}  -  {g}_{2}^{a b}  X_{2}^{\alpha \beta} (u_{R \alpha}^a u_{R \beta}^b) \nonumber\\
& & + \ \lambda X_{1 \alpha} X_{1 \beta} X_{2}^{\alpha \beta} \
\eea
\noindent
Because of the antisymmetric color structure in the first term, $g_1$ must be antisymmetric in flavor which prevents it from introducing meson-antimeson mixing.  The antisymmetric structure of $g_1$ also prevents the existence of six-quark operators involving all first-generation quarks, and thus prevents $n\bar{n}$ oscillations.  As in previous models, $g_2$ is symmetric and so we will get $D^0$-$\bar{D}^0$ mixing as in model 3.  Although this model does not have $n \bar{n}$ oscillations, there are still baryon number violating processes which would constrain this model -- for example, the process $pp \rightarrow K^+ K^+$.  This has been searched using the Super-Kamiokande detector looking for the nucleus decay $^{16} O \rightarrow\, ^{14} C K^+ K^+$ \cite{Litos:2010}.  Had we included $\nu_R$, model 4 would have been excluded by tree-level scalar exchange.

Now, a similar line of reasoning applies to the case where we have a quartic scalar interaction term $X_1 X_1 X_1 X_2$.  The only models violating baryon number which don't generate proton decay (or bound neutron decay) are discussed briefly below.  These last five models have dinucleon decay to leptons, but don't contribute to tree-level $n\bar{n}$ oscillations by virtue of their coupling to leptons.
\\

\noindent $\textbf{Model 5.\ \ \ }$ $X_{1} \in \left(\bar{6}, 1, -1/3\right)$, \ $X_{2} \in \left(1, 1, 1\right)$
\bea
\mathcal{L} & = &  - \ g_{1}^{a b} X_{1}^{\alpha \beta} \left(Q_{L \alpha}^a \epsilon \,Q_{L \beta}^b\right)- {g}_{2}^{a b}  X_{2} (L_{L}^a \epsilon L_{L}^b)\nonumber\\
& & - \ {g}_{1}^{\prime a b}  X_{1}^{\alpha \beta} (u_{R \alpha}^a d_{R \beta}^b) \nonumber\\
 & & + \ \lambda X_{1}^{\alpha \alpha'} X_{1}^{\beta \beta'} X_1^{\gamma \gamma'} X_2\, \epsilon_{\alpha \beta \gamma} \epsilon_{\alpha'\beta'\gamma'}\
\eea
\noindent
Similar arguments to those for the previous models tell us that $g_1$ and $g_2$ must be antisymmetric in flavor.
\\ \\
\noindent $\textbf{Model 6.\ \ \ }$ $X_{1} \in \left(\bar{6}, 3, -1/3\right)$, \ $X_{2} \in \left(1, 1, 1\right)$
\bea
\mathcal{L} & = &  - \ g_{1}^{a b} X_{1}^{\alpha \beta A} (Q_{L \alpha}^a \epsilon \,\tau^A \,Q_{L \beta}^b) - \ {g}_{2}^{a b}  X_{2} (L_{L}^a \epsilon L_{L}^b) \nonumber\\
& &  + \ \lambda X_{1}^{\alpha \alpha' A} X_{1}^{\beta \beta' B} X_1^{\gamma \gamma' C} X_2\, \epsilon^{A B C} \epsilon_{\alpha \beta \gamma} \epsilon_{\alpha'\beta'\gamma'} \
\eea
\noindent
By comparison with model 2, we see that $g_1$ is symmetric in flavor while $g_2$ is antisymmetric.
\\ \\
\noindent $\textbf{Model 7.\ \ \ }$  $X_{1} \in \left(\bar{6}, 3, -1/3\right)$, \ $X_{2} \in \left(1, 3, 1\right)$
\bea
\mathcal{L} & \!\!=\!\! &  - \ g_{1}^{a b} X_{1}^{\alpha \beta A} (Q_{L \alpha}^a \epsilon \,\tau^A \,Q_{L \beta}^b) -  {g}_{2}^{a b}  X_{2}^A (L_{L}^a \epsilon \tau^A L_{L}^b) \nonumber\\
& &  + \ \lambda X_{1}^{\alpha \alpha' A} X_{1}^{\beta \beta' B} X_1^{\gamma \gamma' C} X_2^D\,
\epsilon_{\alpha \beta \gamma}  \epsilon_{\alpha'\beta'\gamma'}  \nonumber\\
& & \times \ (\delta^{AB} \delta^{CD}+\delta^{AC} \delta^{BD}+\delta^{AD} \delta^{BC}) \
\eea
\noindent
Once again, as in model 2, we have a symmetric $g_1$.  The coupling $g_2$ must be symmetric in flavor as well.
\\ \\
\noindent $\textbf{Model 8.\ \ \ }$ $X_{1} \in \left(\bar{6}, 1, 2/3\right)$, \ $X_{2} \in \left(1, 1, -2\right)$
\bea
\mathcal{L} & = &  - \ g_{1}^{a b} X_{1}^{\alpha \beta} (d_{R \alpha}^a d_{R \beta}^b)
-g_{2}^{a b} X_{2} (e_R^a e_R^b) \nonumber\\
& & + \ \lambda X_{1}^{\alpha \alpha'} X_{1}^{\beta \beta'} X_{1}^{\gamma \gamma'} X_2 \epsilon_{\alpha \beta \gamma} \, \epsilon_{\alpha'\beta'\gamma'}\
\eea
\noindent
As in model 1, $g_1$ must be symmetric. The coupling $g_2$ must also be symmetric in flavor.
\\ \\
\noindent $\textbf{Model 9.\ \ \ }$ $X_{1} \in \left(3, 1, 2/3\right)$, \ $X_{2} \in \left(1, 1, -2\right)$
\bea
\mathcal{L} & = &  - \ g_{1}^{a b} X_{1 \alpha} (d_{R \beta}^a d_{R \gamma}^b) \epsilon^{\alpha\beta\gamma}
-g_{2}^{a b} X_{2} (e_R^a e_R^b) \nonumber\\
& & + \ \lambda X_{1 \alpha} X_{1 \beta} X_{1 \gamma} X_{2} \, \epsilon^{\alpha\beta\gamma}\
\eea
\noindent
By comparison with model 4, we see that $g_1$ must be antisymmetric in flavor. The coupling $g_2$ is symmetric.  Note that the antisymmetric color structure of the scalar interaction requires the existence of at least three different kinds of $X_1$ scalars for this coupling to exist.  Including $\nu_R$ would eliminate model 9 for the same reason as model 4.

\section{Phenomenology of model 1}
In this section we present a detailed analysis of model 1. The corresponding calculations for the other models can be performed in a similar manner. Our work is partly motivated by the recently proposed $n\bar{n}$ oscillation experiment with increased sensitivity \cite{projectX}. In addition to $n\bar{n}$ oscillations, we analyze also the cosmological baryon asymmetry generation in model 1 and comment on its LHC phenomenology, as well as flavor and electric dipole moment constraints.

\subsection{Neutron-antineutron oscillations}
The topic of $n\bar{n}$ oscillations has been explored in the literature in various contexts. For some of the early works on the subject see \cite{Kuo:1980ew,Mohapatra:1980qe,Ozer:1982qh}. Recently, a preliminary study of the required hadronic matrix elements using  lattice QCD has been carried out \cite{Buchoff:2012bm}. Reference \cite{Berezhiani:2012rq} claims that a signal of $n\bar{n}$ oscillations has been observed.

The scalar content of model 1 we are considering is similar to the content of a unified model explored in \cite{Babu:2012vc}.  The transition matrix element,
\bea
\Delta m = \langle \bar{n}| \mathcal{H}_{\rm eff} |n\rangle\ ,
\label{Heff}
\eea
leads to a transition probability for a neutron at rest to change into an antineutron after time $t$ equal to $P_{n\rightarrow \bar{n}}(t) = \sin^2\!\left(|\Delta m |\, t\right)$.

\begin{figure}[t!]
\centering \hspace{5mm}
\includegraphics[scale=.65]{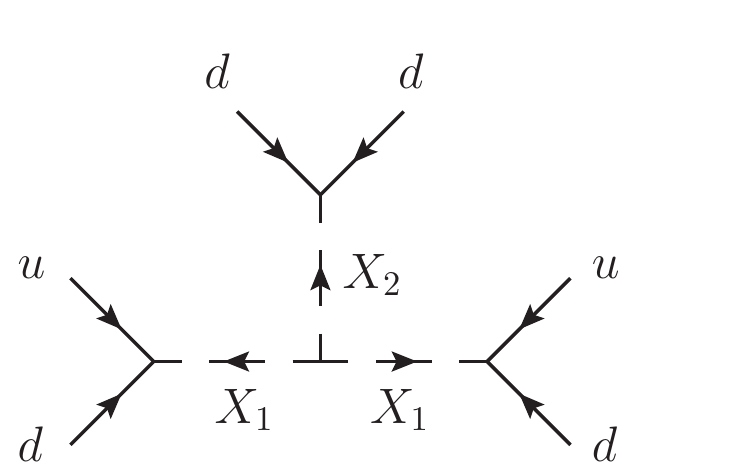}
\caption{Interaction which leads to neutron-antineutron oscillations.}
\label{fig1b}
\end{figure}

Neglecting the coupling $g_1$ in the Lagrangian (\ref{Lag1}) (for simplicity) the effective $|\Delta B|=2$ Hamiltonian that causes $n\bar{n}$ oscillations is,
\bea
\mathcal{H_{\rm eff}} &\!\!\!=\!\!\!& - \frac{(g_1^{\prime 11})^2 g_2^{11} \lambda}{4 M_{1}^4 M_{2}^2} d_{R i}^{\dot{\alpha}} d_{R i'}^{\dot{\beta}} u_{R j}^{\dot{\gamma}} d_{R j'}^{\dot{\delta}} u_{R k}^{\dot{\lambda}} d_{R k'}^{\dot{\chi}} \epsilon_{\dot{\alpha} \dot{\beta}} \epsilon_{\dot{\gamma} \dot{\delta}} \epsilon_{\dot{\lambda} \dot{\chi}}\nonumber\\
& &\hspace{-10mm}\times \Big(\epsilon_{ijk} \epsilon_{i'j'k'} + \epsilon_{i'jk} \epsilon_{ij'k'}+\epsilon_{ij'k} \epsilon_{i'j k'}+\epsilon_{ijk'} \epsilon_{i'j'k}\Big) +{\rm h.c.}\nonumber\\
\eea
where Latin indices are color and Greek indices are spinor. It arises from the tree-level diagram in Fig.\ \ref{fig1b}.  We have rotated the couplings $g_1'$ and $g_2$ to the quark mass eigenstate basis and adopted a phase convention where $\lambda$ is real and positive.  We estimate $\Delta m$ using the vacuum insertion approximation \cite{Gaillard:1974hs}.  This relates the required $n\bar{n}$ six quark matrix element to a matrix element from the neutron to the vacuum of a three quark operator.  The later matrix element is relevant for proton decay and has been determined using lattice QCD methods.  The general form of the required hadronic matrix elements is,

\begin{align}
\langle 0 &|d_{R i}^{\dot{\alpha}} d_{R j}^{\dot{\beta}} u_{R k}^{\dot{\gamma}} | n(p,s) \rangle \nonumber\\
 &\ \ \ \ \ \ \ \ \ = -\frac{1}{18}  \beta ~\epsilon_{ijk}
\left(\epsilon^{\dot{\alpha} \dot{\gamma}} u_R^{\dot{\beta}}(p,s)+\epsilon^{\dot{\beta} \dot{\gamma}} u_R^{\dot{\alpha}}(p,s)\right)\ .
\end{align}
Here $u_R$ is the right-handed neutron two-component spinor and the Dirac equation was used to remove the term proportional to the left-handed neutron spinor.  The constant $\beta$ was determined using lattice methods in Ref.\ \cite{Tsutsui:2004qc} to have the value  $\beta \simeq 0.01~{\rm GeV}^3$.   In  the vacuum insertion approximation  to Eq.\ (\ref{Heff}) we find that,
\bea\label{deltam}
|\Delta m |= 2 \lambda \beta^2\frac{|  (g_1^{\prime 11})^2 g_2^{11}| }{3 M_{1}^4 M_{2}^2}\ .
\eea
The current experimental limit on $\Delta m$ is \cite{Abe:2011ky},
\bea
|\Delta m| < 2\times 10^{-33} \ {\rm GeV}\ .
\eea
For scalars of equal mass, $M_1 = M_2 \equiv M$, and the values of the couplings $g_1^{\prime 11} = g_2^{11} = 1$, $\lambda = M$, one obtains,
\bea
M \gtrsim 500 \ {\rm TeV}\ .
\eea
If, instead, the masses form a hierarchy, the effect on $n\bar{n}$ oscillations is maximized if we choose $M_2 > M_1$.   Assuming $M_1 = 5 \ {\rm TeV}$ (above the current LHC reach) and $\lambda = M_2 $ this yields,
\bea
M_2 \gtrsim 5\times 10^{13} \ {\rm GeV}\ .
\eea
Note that $\lambda=M_2$ is a reasonable value for this coupling since integrating out $M_2$ then gives a quartic $X_1$ interaction term with a coupling on the order of one. Of course, this model does have a hierarchy problem so having the Higgs scalar and the $X_1$ light compared with $X_2$ requires fine tuning.

Experiments  in the future \cite{projectX} may be able to probe $n\bar{n}$ oscillations with increased sensitivity of $|\Delta m| \simeq 7\times 10^{-35} \ \rm GeV$. If no oscillations are observed, the new limit in the case of equal  masses will be,
\bea
M \gtrsim  1000 \ {\rm TeV}\ .
\eea
On the other hand, having $M_1 = 5 \ {\rm TeV}$ would push the mass of the heavier scalar up to the GUT scale, leading to the following constraint on the second scalar mass,
\bea
M_2 \gtrsim 1.5\times 10^{15} \ {\rm GeV}\ .
\eea
We note, however, that in Section \ref{flavorsection} we show that $M_1$ on the order of a few TeV is strongly disfavored by the electric dipole moment constraints.

\subsection{LHC, flavor and electric dipole moment constraints}\label{flavorsection}

If the mass of the scalar $X_1$ is small enough, it can be  produced at the LHC through both single and pair production. Detailed analyses have been performed setting limits on the mass of $X_1$ from such processes \cite{Richardson:2011df,Berger:2010fy,Chen:2008hh}. A recent simulation \cite{Richardson:2011df} shows that 100 $\rm fb^{-1}$ of data from the LHC running at 14 TeV center of mass energy can be used to rule out or claim a discovery of $X_1$ scalars with masses only up to approximately 1 TeV, even when the couplings to quarks are of order $1$. Our earlier choice of $M_1 = 5 \ \rm TeV$ used to estimate the constraint on $M_2$ from $n\bar{n}$ oscillations lies well within the allowed mass region.

\begin{figure}[t!]
\includegraphics[height=1.05in]{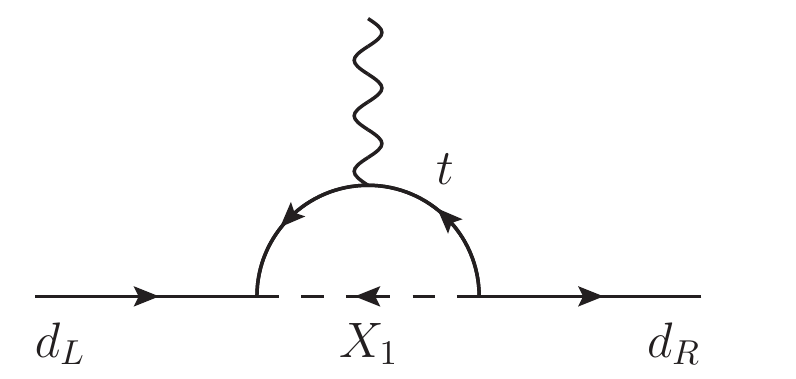}
\caption{Diagrams contributing to the electric dipole moment of the down quark.}
\label{fig5}
\end{figure}

Some of the most stringent flavor constraints on new scalars come from neutral meson mixing and electric dipole moments.
The fact that in model 1 $X_1$ couples directly to both left- and right-handed quarks means that at one loop the top quark mass can induce the chirality flip necessary to give a light quark edm, putting strong constraints on this model even when $X_1$ is at the $100~{\rm TeV}$ scale.  The diagram contributing to the edm of the down quark is given in Fig.\ \ref{fig5}. We find,
\bea\label{dipole2}
\left|d_d\right| \simeq \frac{m_t}{6 \pi^2 M_1^2}\log\left(\frac{M_1^2}{m_t^2}\right)\,\left|{\rm Im}[g_1^{31} ({g'}_1^{31})^*]\right| \ \ e\ \,{\rm cm}\ .
\eea
Here we have neglected pieces not logarithmically enhanced.  This will give the largest contribution to the neutron edm because of the top quark mass factor.  All Yukawa couplings in this section are in the mass eigenstate basis.

Using $SU(6)$ wavefunctions, this can be related to the neutron edm via $d_n = \frac{4}{3} d_d - \frac{1}{3} d_u \simeq \frac{4}{3} d_d$. The present experimental limit is \cite{Baker:2006ts},
\bea
d_n^{\rm exp} < 2.9 \times 10^{-26} \ \ e\ \,{\rm cm}\ .
\eea
Assuming $M_1 = 500 \ {\rm TeV}$, neutron edm measurements imply the bound $\big|{\rm Im}[g_1^{31} ({g'}_1^{31})^*]\big| \lesssim 6\times 10^{-3}$. Furthermore, for observable $n\bar{n}$ oscillation effects with $M_2$ being close to the GUT scale we need $M_1 \approx 5 \ {\rm TeV}$. In such a scenario the edm constraint requires $\big|{\rm Im}[g_1^{31} ({g'}_1^{31})^*]\big| \lesssim  10^{-6}$.

Another important constraint on the parameters of model 1 is provided by $K^0$-$\bar{K}^0$ mixing.  Integrating out $X_2$ generates an effective Hamiltonian,
\begin{align}
\mathcal{H}_{\rm eff}=&\frac{g_2^{22} \left(g_2^{11}\right)^*}{M_2^2} (s_{R \alpha} s_{R \beta})(d_R^{* \alpha} s_R^{* \beta})\nonumber \\
\rightarrow ~ &\frac{g_2^{22} \left(g_2^{11}\right)^*}{2 M_2^2} (\bar{d}_R^{\alpha} \gamma^{\mu} s_{R \alpha}) (\bar{d}_R^{\alpha} \gamma_{\mu} s_{R \alpha}) ,
\end{align}
where in the second line we have gone from two- to four-component spinor notation.  This gives the following constraints on the couplings \cite{Isidori:2010kg},
\begin{align}
\left| {\rm Re}\!\left[ g_2^{22} \left(g_2^{11}\right)^* \right]\right| &<  1.8 \times 10^{-6} \left( \frac{M_2}{1 ~ {\rm TeV}} \right)^2 , \\
\left| {\rm Im}\!\left[ g_2^{22} \left(g_2^{11}\right)^* \right]\right| &<  6.8 \times 10^{-9} \left( \frac{M_2}{1 ~ {\rm TeV}} \right)^2 .
\end{align}
If we set $M_2$ to 500 TeV, this corresponds to an upper bound on the real and imaginary parts of $g_2^{22} \left(g_2^{11}\right)^*$ of 0.45 and $1.7 \times 10^{-3}$, respectively.

\subsection{Baryon asymmetry}
\begin{figure}[t]
$\begin{array}{cc}
\includegraphics[height=1.05in]{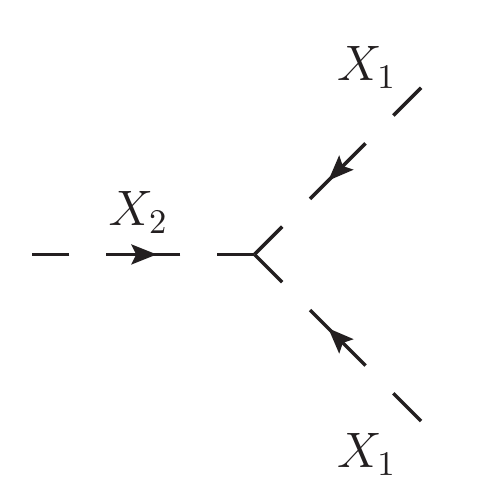} & \includegraphics[height=1.05in]{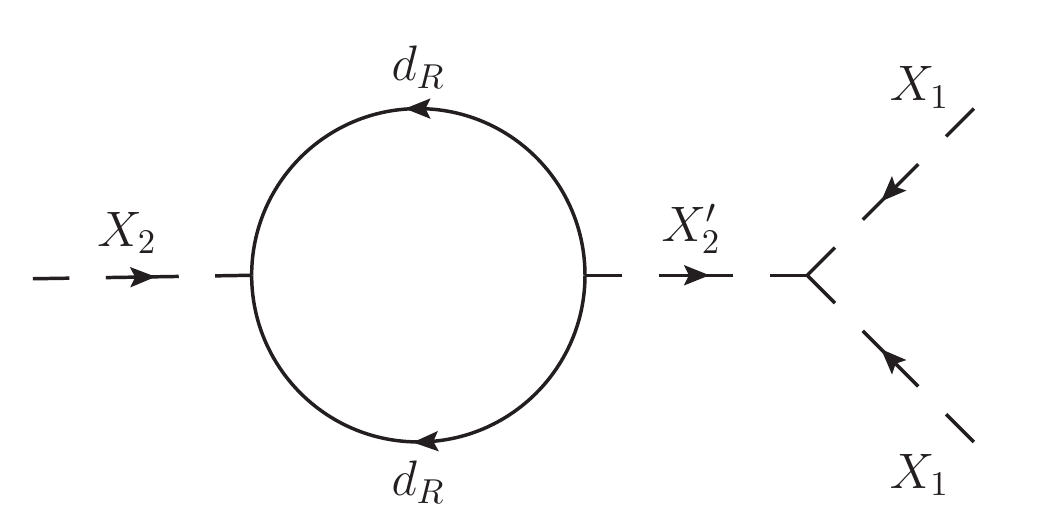} \\
\includegraphics[height=1.05in]{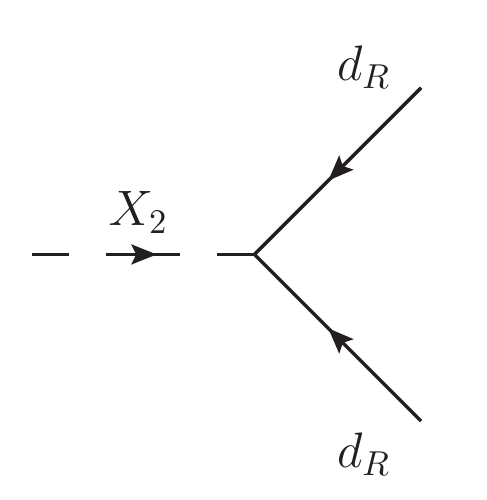} & \includegraphics[height=1.05in]{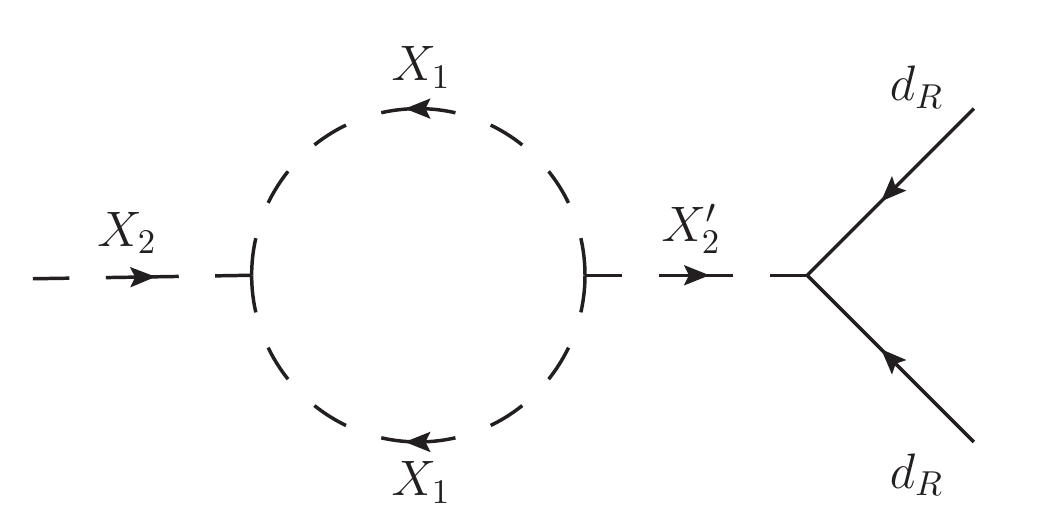}
\end{array}$
\caption{Diagrams corresponding to the decay of $X_2$.  The diagrams on top contribute to the $\Delta B = 2$ decays, while the diagrams on bottom contribute to $\Delta B = 0$.}
\label{fig6}\end{figure}
\begin{table}[b]
\begin{tabular}{|c|c|c|}
\hline
~~~~~~~~ Decay ~~~~~~~~&~~~~ Br ~~~~&~~ $B_f$ ~~ \\ \hline \hline
$X_2 \rightarrow \overline{X}_1 \overline{X}_1$ & $r$ & $4/3$\\ \hline
$X_2 \rightarrow \bar{d}_R \bar{d}_R$ & $1-r$ & $-2/3$\\ \hline
$\overline{X}_2 \rightarrow X_1 X_1$ & $\bar{r}$ & $-4/3$\\ \hline
$\overline{X}_2 \rightarrow d_R d_R$ & $1-\bar{r}$ & $2/3$\\ \hline
\end{tabular}
\caption{Branching ratios and final state baryon numbers for the decays of $X_2$ and $\overline{X}_2$ which contribute to the baryon asymmetry in the coupling hierarchy $\lambda, \tilde{\lambda} \ll g_2, \tilde{g}_2$.}\label{table2}
\end{table}
We now investigate baryon number generation in model 1. $B$ and $L$ violating processes in cosmology have been studied in the literature in great detail (for early works, see \cite{Nanopoulos:1979gx,Harvey:1990qw}).
We treat $X_2$ as much heavier than $X_1$ and use two different $X_2$'s to get a CP violating phase in the one-loop diagrams that generate the baryon asymmetry. For this calculation $X_1$ is treated as stable with baryon number -2/3 as each will eventually decay via baryon number conserving processes to two antiquarks.  To simplify our discussion, let's consider the case in which the couplings satisfy the hierarchy $\lambda, \tilde{\lambda} \ll g_2, \tilde{g}_2$. The top line of Fig.\ \ref{fig6} shows the dominant tree-level and one-loop diagrams contributing to the baryon number violating decays of $X_2$.  Rotating the $X$ fields to make the couplings $\lambda$ and ${\tilde \lambda}$ real we find,
\bea
& & \hspace{-6mm}\Gamma(X_2 \rightarrow \overline{X}_1 \overline{X}_1) = \frac{3 \lambda}{8\pi M_{2}} \nonumber\\
&& \ \ \ \ \ \hspace{-2mm}\times\left[\lambda-{\tilde \lambda}\frac{M_{2}^2}{4\pi (M_{2}^2-{\tilde{M}_{2}}^2)}{\rm Im}({\rm Tr}(g_{2 }^\dagger\, \tilde{g}_{2 }))\right]\!,\nonumber\\
& & \hspace{-6mm}\Gamma(\overline{X}_2 \rightarrow X_1 X_1) = \frac{3 \lambda}{8\pi M_{2}}\nonumber\\
&& \ \ \ \ \ \hspace{-2mm}\times\left[\lambda+{\tilde \lambda}\frac{M_{2}^2}{4\pi (M_{2}^2-{\tilde{M}_{2}}^2)}{\rm Im}({\rm Tr}(g_{2 }^\dagger\, \tilde{g}_{2 }))\right]\!.\eea
The net baryon number produced per $X_2\overline{X}_2$  pair  is (see, Table \ref{table2}),
\begin{align}\label{delta_n}
\Delta n_B &= 2 (r-\bar{r})\nonumber \\
&= \frac{6}{\pi \,{\rm Tr}(g_2^{\dagger} g_2)} \frac{1}{\tilde{M}_2^2-M_2^2}\,{\rm Im}\!\left[\lambda \,\tilde{\lambda}^{*}\,{\rm Tr}(g_{2 }^\dagger\, \tilde{g}_{2})\right],
\end{align}
where we have used the fact that $CPT$ invariance guarantees the total width of $X_2$ and $\bar{X}_2$ are the same. Given our choice of hierarchy for the couplings, we have approximated the total width as coming from the tree-level decay of $X_2$ to antiquarks.
A similar result in the context of $SO(10)$ models was obtained in Ref.\ \cite{Babu:2012vc}.

Even with just one generation of quarks, the CP violating phase cannot be removed from the couplings $ \lambda$, ${\tilde \lambda}$, $g_2$, ${\tilde g_2}$ and a baryon asymmetry can be generated at one loop. At first glance this is surprising since there are four fields, $X_2,$ ${\tilde X}_2$, $X_1$ and $d_R$ whose phases can be redefined and four relevant couplings. However, this can be understood by looking at the relevant Lagrangian terms, $g_2 X_2 d d$, $\tilde{g}_2 \tilde{X}_2 d d$, $\lambda X_1 X_1 X_2$ and $\tilde{\lambda} X_1 X_1 \tilde{X_2}$. The problem reduces to finding solutions to the following matrix equation,
\bea
\left(
  \begin{array}{cccc}
    2 & 1 & 0 & 0 \\
    2 & 0 & 1 & 0 \\
    0 & 1 & 0 & 2 \\
    0 & 0 & 1 & 2 \\
  \end{array}
  \right)\left(
         \begin{array}{c}
           \phi_{X_1} \\
           \phi_{X_2} \\
           \phi_{{\tilde X}_2} \\
           \phi_d \\
         \end{array}
       \right) = \left(
         \begin{array}{c}
           \phi_{\lambda} \\
                      \phi_{\tilde{\lambda}} \\ \phi_{g_2} \\
                                 \phi_{{\tilde g}_2} \\
         \end{array}
       \right)\ ,
\eea
where the phases on the right-hand side are arbitrary. Let us take the difference of the first two equations to remove phases for the couplings $\lambda$ and $\tilde{\lambda}$, and the difference of the last two equations to remove phases for the coupling $g_2, \tilde{g}_2$. We therefore obtain $\phi_{{\tilde \lambda}_2}-\phi_{{\lambda}_2} =\phi_{\tilde{X}_2}-\phi_{{X}_2} $ and $\phi_{{\tilde g}_2}-\phi_{{g}_2} =\phi_{\tilde{X}_2}-\phi_{{X}_2}$. Those two equations cannot be simultaneously fulfilled for arbitrary $ \phi_\lambda$, $\phi_{\tilde \lambda}$, $\phi_{g_2}$, $\phi_{\tilde g_2}$.

The baryon number generated in the early universe can be calculated from Eq.\ (\ref{delta_n}) by following the usual steps (see, for example, \cite{Babu:2012vb}).  Out of equilibrium decay of $X_2$ and $\bar{X}_2$ is most plausible if they are very heavy (e.g.\ $\sim 10^{12} ~{\rm GeV}$).  However, to get measurable $n\bar{n}$ oscillation in this case, $X_1$ would have to be light -- a case that is strongly disfavored by neutron edm constraints.
\\
\\
\section{Conclusions}
We have investigated a set of minimal models which violate baryon number at tree-level without inducing proton decay.  We have looked in detail at the phenomenological aspects of one of these models (model 1) which can have $n\bar{n}$ oscillations within the reach of future experiments. When all the mass parameters in model 1 have the same value, $M$, and the magnitudes of the Yukawa couplings $g_1^{\prime 1 1}$ and $g_2^{1 1}$ are unity, the present limit on $n \bar{n}$ oscillations implies that $M$ is greater than $500 ~{\rm TeV}$.  For $M=500 ~{\rm TeV}$, the neutron edm and flavor constraints give ${\rm Im}\! \left[g_1^{31} (g_1^{\prime 31})^* \right] < 6 \times 10^{-6}$, ${\rm Re}\! \left[g_2^{22} (g_2^{11})^* \right] < 0.45$, and ${\rm Im}\! \left[g_2^{22} (g_2^{11})^* \right] < 1.7 \times 10^{-3}$ which indicates that some of the Yukawa couplings and/or their phases must be small if $n\bar{n}$ oscillations are to be observed in the next generation of experiments. Of course even in the standard model some of the Yukawa couplings are small.

There are two other models (model 2 and model 3) that have $n \bar{n}$ oscillations at tree-level.  Similar conclusions can be drawn for them, although the details are different.  In models 2 and 3, exchange of a single $X_1$ does not give rise to a one-loop edm of the neutron.  However, $K^0$-$ \bar{K}^0$ mixing can occur from tree-level $X_1$ exchange.

Observable $n \bar{n}$ oscillations can occur for $M_2 \gg M_1$ with $M_2$ at/near the GUT scale.  This requires $M_1 \simeq 5~{\rm TeV}$, and flavor and electric dipole constraints require some very small Yukawa couplings making this scenario unlikely.

\subsection*{Acknowledgment}
The work of the authors was supported in part by the U.S. Department of Energy under contract No. DE-FG02-92ER40701 and by the Gordon and Betty Moore Foundation.  We thank P.\ F.\ P\'{e}rez for helpful comments.


\end{document}